# Ordered Short-Range Ripple Effects in Structures of Silicenes: Role of Puckering in the Aromatic Rings

Ayan Datta

*School of Chemistry, Indian Institute of Science Education and Research Thiruvananthapuram, CET Campus, Thiruvananthapuram-695016, Kerala (India)*

**Structural and electronic properties of the all-Si analogue of graphene, silicene have elucidated through DFT calculations. Silicene differs considerably from graphene in being 'chair-type' puckered in each 6-membered ring which leads to ordered ripples across the surface. Binding energies suggest stability for such rippled silicenes and are predicted to behave as a finite gap semi-conductor with electron-hole symmetry quenched. Inter-layer coupling between the silicenes is suggested as the mechanism for the formation of the bulk-Si in it's only known diamond form.**

Graphene has attracted immense interest in the present decade due to its remarkable chemical, physical, mechanical, electronic and magnetic properties.[1-3] This nanoscale 2-D system provides a wonderful starting material for fabricating materials in the nano dimension. Apart from its novel prospects, the fundamental structural aspects of graphene are also very interesting.[4-5] Though expected to be ideally planar, recent experiments and computations suggest that graphene is not really planar and distinct ripples are observed.[6-8] This is also expected from the Mermin-Wagner theorem predicting that thermal fluctuations should destroy long-range order in 2-D systems and instabilities should appear. This phenomenon is similar to the established Peierls distortion in polyacetylene and other 1-D systems.[9]

The all silicon analogue of graphene, silicene has generated recent interest. Silicon nanoribbons have been studied through STM studies and DFT calculations.[10-11] Even though there is an immediate possibility of application of silicene based nano-materials in existing Si-microelectronics, the fundamental structural aspects are yet to be elucidated. In this communication, we report the rich structural and electronic aspects in nanosheets of polysilo-acenes and silicenes based on DFT calculations. In marked contrast to graphene, the ground-state of silicene show large, short-range and periodically ordered ripples even in the absence of thermal fluctuations. This effect is a direct consequence of the puckering distortion in the six-membered rings that leads to a gap opening in silicene, unlike the zero-gap semiconductor like behavior in graphene.

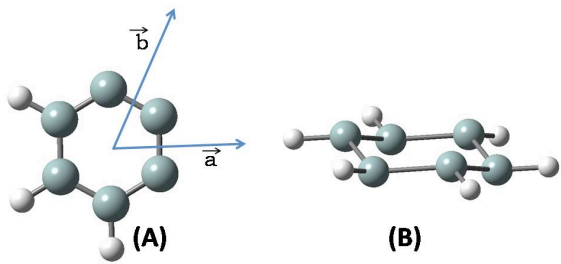

**Figure 1.** (A) The translational units used to construct silicenes of various nuclearities. (B) Structure of a single aromatic unit of silicene, hexasilabenzene.

Fig. 1 (A) shows the translational units on which our calculations are performed. The silicene fragments are constructed using the units (m,n) along the vectors a and b. (m,n) are varied from (1,1) to (5,5). The valency for the terminal Si-atoms were satisfied through Si-H bonds. Thus, we have studied silicene clusters from a 1 six-membered ring (hexasilabenzene) to a 25 ring cluster. The structures for all the clusters were optimized using the hybrid B3PW91 DFT functional[12] at the triple-ξ-valence polarized basis set level[13-14]. Additional frequency calculations were performed for ensuring the absence of any vibrational instabilities in the structures. A general feature for all the silicene clusters is the $C_3$ puckering distortion in the six-membered rings. This is clearly seen in Fig. 1(B) for hexasilabenzene in harmony with previous experimental and computational results.[15-16] The dihedral angle for puckering in hexasilabenzene is 33.69 degrees. Interestingly, this dihedral angle remains almost constant for all values of (m,n) [varying between 33 degrees to 36 degrees] (see supporting Info. File for structures). The high puckering symmetry of the silicene clusters ensure that the puckering in each ring leads to an ordered ripple across the surfaces. The ripple is localized within the six-membered ring and fresh ripples appear in the adjoining rings. In Fig. 2 (A), the ripples for a typical silicene cluster are clearly seen.

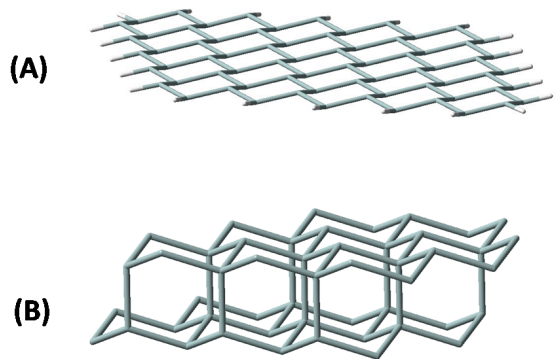

**Figure 2.** (A) Ripples in (5,5) silicene strip. (B) Structure of bulk Si viewed along the rippled layers.

The stability of the silicene strips is estimated by calculating the binding energy of the strip, $\Delta E = E(Si_aH_b) - (a/2)E(Si_2H_2) + [(a-b)/2]E(H_2)$, where a and b are number of Si an H atoms in the silicene strip respectively. The presence of six-membered rings as the basic building blocks in silicenes and choice of disilyne[17] as the fractional unit ensure that both a and b are even integers. In Fig. 3, the profile for variation in the binding energy per Si-atom of the silicene strips for various m and n is shown. It is interesting to note that even though for small strips (n,m ≤ 2) the binding energies show odd-even oscillations as a consequence of finite-size effects, the binding energies start to converge for larger strips.

The binding energy per Si-atom is -1.026 eV for (5,5) silicene cluster and an asymptotic fitting leads to a value -1.21 eV for the infinite strip. It is important to note that such formidable binding energy suggest stability and existence of silicene even in the absence of a surface support (Basis set superposition error might contributed ~ 10% to the stabilization energies without changing the overall conclusions). Also, favorable binding energies for the small silicene fragments like $Si_{10}H_8$, $Si_{14}H_{10}$, $Si_{16}H_{10}$, $Si_{18}H_{12}$ and $Si_{22}H_{14}$ opens an interesting possibility for the existence of all-Si analogues of naphthalene, anthracene, perylene, tetracene and pentacene respectively.

The puckering distortion in hexasilabenzene and the consequent ripples in silicene arises from the poor overlap between the $3p_z$ orbitals on Si-atoms which leads to the poor electron-electron interactions being overcome by the electron-nuclear interactions and consequent deviation from planarity. This is opposite to the case of graphene. Interestingly, calculations that do not consider such interactions predict a planar graphene – like structure for silicene and similar electronic properties.[18] However, a detailed search in the potential energy surface (PES) for silicenes predict a puckered structure in the six – membered rings which supports our calculations.[19]

Such, ripples thus are expected to strongly affect the band-gap of materials. To understand the consequences of such distortions in the six – membered rings, we have performed additional periodic DFT calculations on the puckred silicene layers. In Fig. 4, we plot the density-of-states (DOS) for silicene using periodic boundary conditions for the extended puckered layers in 2-D at PW91/TZP level[20]. Silicene is calculated as a finite (0.89 eV) band-gap semi-conductor with asymmetric band-structure states above and below the Fermi-Energy ($E_F$). Such a break-down of the electron-hole symmetry is a direct consequence of the ordered ripples in silicene. Interestingly, a planar 2D-silicene is expected to be metallic.[19]

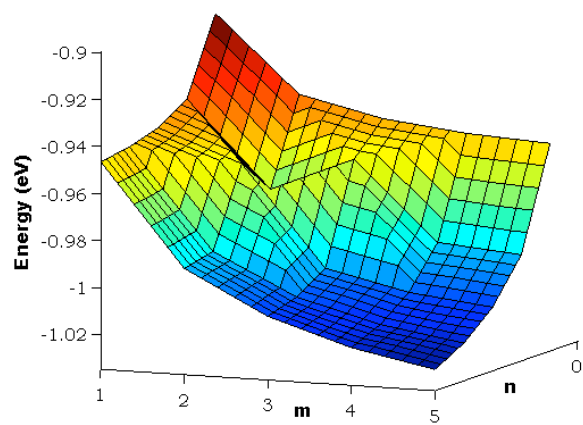

**Figure 3.** Binding energy per Si-atom (in eV) for silicenes of various nuclearilites.

It is interesting to note that in the bulk, Si is known to exist in only one form with a FCC diamond cubic structure[21] which might be visualized as layers of puckered silicenes stacked over one another with the $C_3$ distortions arising out of the additional interlayer Si-Si bonds (Fig. 2 (B)). This is in contrast to carbon that additionally exists in the pure $sp^2$ graphite structure. Thus, the absence of any other allotrope of Si similar to the graphitic form in carbon can be attributed to the puckering distortion that renders π-stacking interactions ineffective through the loss of planarity in each six membered ring and favors an $sp^3$ bonding environment. Also, the diamond structure of Si ensures that Si-Si interaction energies are identical both along the layer and in-between the layers.

In conclusion, we have predicted in this communication that the 2-D sheet of all-Si structure, silicene should exist independently without supporting matrix. Interestingly, recent calculations have shown that silicene clusters can be stabilized through complexation to transition – metals to form closed shell 18 – electron complexes.[22]

However, such layers exhibit unique ordered ripples across the surface arising out of 'chair-type' distortions in each six membered rings. Hence, ripples lead to opening up of a gap in the band-structure and suggests impressive electronic applications through structural designing. These ripples in silicene layers also explain the formation of bulk Si through interlayer coupling. It would be interesting to grow crystals of all-Si analogues of tetracene and pentacene as materials for electron/hole transport in the mesoscopic scale.

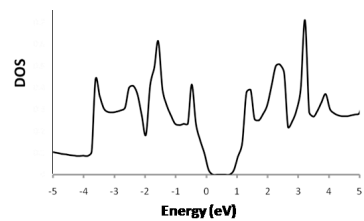

**Figure 4.** Density-of-States (DOS) for silicene.

**Acknowledgement** AD thanks DST-fast Track Scheme (Govt. of India) for partial research funding.

**Supporting Information Available:**. Cartesian coordinates, Energies, harmonic frequencies for silicenes of various nuclearities and complete reference 14. This material is available free of charge through Internet.